\documentclass[3p,times]{elsarticle}

\usepackage{ecrc}
\usepackage{floatrow}

\volume{00}

\firstpage{1}

\journalname{Nuclear Instruments and Methods A}

\runauth{J.I. Collar et al.}

\jid{pdu}

\jnltitlelogo{NIM A}

\usepackage{amssymb}

\usepackage[figuresright]{rotating}

\begin{document}

\begin{frontmatter}



\dochead{}

\title{Coherent neutrino-nucleus scattering detection with a CsI[Na] scintillator at the SNS  spallation source}


\ead{collar@uchicago.edu}

\author{J.I.~Collar$^{1}$, N.E.~Fields$^{1}$, M.~Hai$^{1,\dag}$, T.W.~Hossbach$^{2}$, J.L.~Orrell$^{2}$, C.T.~Overman$^{2}$,G.~Perumpilly$^{1}$, B.~Scholz$^{1}$}

\address{$^{1}$Enrico Fermi Institute, Kavli Institute for 
Cosmological Physics, and Department of Physics,
University of Chicago, Chicago, IL 60637, USA}

\address{$^{2}$Pacific Northwest National Laboratory, Richland, WA 99352, USA}

\address{$^{\dag}$Present address: College of Engineering, Carnegie Mellon University, Pittsburgh, PA 15213, USA}

\begin{abstract}
We study the possibility of using CsI[Na] scintillators as an advantageous target for the detection of coherent elastic neutrino-nucleus scattering (CENNS), using the neutrino emissions from the SNS spallation source at Oak Ridge National Laboratory.  The response of this material to low-energy nuclear recoils like those expected from this process is characterized. Backgrounds are studied using a 2 kg low-background prototype crystal in a dedicated radiation shield. The conclusion is that a planned 14 kg detector should measure approximately 550 CENNS events per year above a demonstrated $\sim7$ keVnr low-energy threshold, with a signal-to-background ratio sufficient for a first measurement of the CENNS cross-section. The cross-section for the $^{208}$Pb($\nu_{e},e^{-}$)$^{208}$Bi reaction, of interest for future supernova neutrino detection, can be simultaneously obtained.
\end{abstract}

\begin{keyword}
neutrino interactions \sep spallation sources \sep cesium iodide \sep coherent neutrino scattering \sep CENNS

\end{keyword}

\end{frontmatter}

\section{Introduction}
\label{}

Neutrinos with energies below few tens of MeV are expected to undergo a large enhancement to the cross-section for their elastic scattering off heavy nuclei \cite{freedman,drukier}. During this neutral-current interaction, all nucleons are expected to contribute coherently to the scattering process\footnote{The contribution from protons is markedly reduced by the numerical value of the weak mixing angle, resulting in a CENNS cross-section effectively proportional to the square of the number of neutrons in the target nucleus \cite{drukier}.}. Coherent elastic neutrino-nucleus scattering (CENNS) is expected to be the dominant mechanism in neutrino transport within supernovae and neutron stars \cite{wilson}, but it remains undetected forty years after its first description. This somewhat puzzling absence of experimental evidence for the largest low-energy neutrino cross-section is due to the modest energy of the nuclear recoils induced, and to the limited intensity of available neutrino sources in this energy range. The coherent nuclear mechanism behind this mode of neutrino interaction is assumed to be taking place also during the spin-independent scattering of favored dark matter candidates (WIMPs), adding to the interest of its experimental verification. Unfulfilled proposals for the detection of CENNS abound \cite{drukier,ieee,steve,ppc,red}.\\

The development of smallish coherent neutrino detectors may eventually enable technological applications \cite{leocastle}. An example is the non-intrusive monitoring of nuclear reactors using compact neutrino detectors, improving on existing methods \cite{bernstein1}. A number of tests for new physics should also be possible with a CENNS-sensitive neutrino detector. For instance, a neutral-current detector responds almost identically to all known neutrino types \cite{all}: observation of neutrino oscillations in such a device 
would provide unequivocal direct evidence for sterile neutrino(s) \cite{drukier}. In addition to this, the differential cross section for this process is strongly dependent on a finite value of the neutrino magnetic moment \cite{dodd}. Recent studies have described the sensitivity of CENNS to non-standard neutrino interactions with quarks \cite{new}, to the effective neutrino charge radius \cite{bernabeu}, and to neutron density distributions in the nucleus \cite{patton}. A precise measurement of the CENNS cross section
would also provide a sensitive appraisal of the weak nuclear charge \cite{larry}.\\

In a seminal paper \cite{drukier}, Drukier and Stodolsky examined the prospects for CENNS detection from a variety of low-energy neutrino sources (solar, terrestrial, supernova, reactor, and spallation source). The opportunity provided by the high-luminosity SNS spallation source at Oak Ridge National Laboratory has been treated in \cite{frankyuri,kate}. While the main use for the SNS is as an intense neutron source, its protons-on-target (POT) produce pions, which decay at rest generating a monochromatic flash of 30 MeV prompt $\nu_{\mu}$. This is followed by delayed  $\nu_{e}$ and $\bar{\nu}_{\mu}$ emissions with a broad energy (Michel spectrum, few tens of MeV), over the $\sim$2.2 $\mu$s timescale characteristic of $\mu$ decay. For present running conditions, the neutrino flux at 20 m from the SNS source is 1.7$\cdot10^{7}$ cm$^{-2}$ s$^{-1}$, per flavor. An attractive feature of this intense neutrino source is its time structure, consisting of 60 Hz of sub-$\mu$s pulses. This provides a $\sim6\cdot10^{-4}$ background rejection factor, when looking for CENNS signals in a sensitive detector, using only the $10~\mu$s window following POT. As will be discussed in Sec.\ 5, unfortunately most POT-associated backgrounds are expected to reach a CENNS detector with a time profile similar to that from the sought neutrino signals. It is therefore of the utmost importance to understand and abate these beam-associated backgrounds. Steady-state backgrounds unrelated to the SNS beam can be characterized during time periods preceding the POT signal. In the absence of beam-related backgrounds, the residual obtained by subtracting pre-POT from post-POT energy spectra is expected to display a characteristic CENNS-induced excess (Sec.\ 5).

\section{Advantages of CsI[Na] as a CENNS detector}
\label{}

 The inorganic CsI[Na] scintillator presents several important advantages for CENNS detection at a spallation source. These can be listed:

\begin{itemize}

\item The high mass of both recoiling species, cesium and iodine, provides a very large coherent enhancement to the CENNS cross-section. This results into a substantial signal event rate (tens of events per kg per year, 20 m from the SNS source), as long as a sufficiently-low detection threshold can be achieved (few keV of nuclear recoil energy, keVnr, Fig.\ 1). 

\item Both recoiling species are essentially indistinguishable due to their very similar mass (Fig.\ 2), greatly simplifying the understanding of the response of the detector. This is of particular importance in the presence of competing backgrounds, like those described in Sec.\ 5. 

\item The light yield from low-energy nuclear recoils in CsI[Na] is sufficiently large to expect a realistic $\sim$7 keVnr threshold when employing conventional bialkali photomultipliers (PMTs) in combination with this scintillator (Sec.\ 3). In the near future, this threshold can be further reduced down to $\sim$4.5 keVnr with the adoption of ultra-bialkali (UBA) photocathodes \cite{UBA}, once this relatively new technology has been extended to large surface area PMTs. A 7 keVnr threshold is demonstrated here, under background conditions similar to those expected at the SNS (Sec.\ 5).  

\item It may be possible to perform statistical discrimination (as opposed to event-by-event discrimination) between low-energy nuclear and electron recoils for $\gtrsim$1,000 accumulated events. This is based on a difference of $\sim$60 ns between the fast scintillation decay time noticed for these two families of events (Sec.\ 3). Similar features have been observed and exploited in NaI[Tl] scintillators dedicated to WIMP detection \cite{nai60}. 

\item CsI[Na] crystals are naturally low in internal radioactivity. When grown from screened salts confirmed to contain low levels of U, Th, $^{40}$K, $^{87}$Rb, and $^{134,137}$Cs \cite{radiopurekims}, as is done here, they provide a sufficient CENNS signal-to-background ratio. A steady-state background subtraction made possible by the pulsed nature of the CENNS signal at the SNS further improves this situation (Sec.\ 5).

\item A convenient low-energy signal at 57.6 keV arises from the inelastic scattering of fast neutrons, $^{127}$I(n,n'$\gamma$). This signal can be monitored to provide an {\it in situ} measurement of neutron backgrounds competing with the CENNS signal (Sec.\ 5).

\item CsI[Na] lacks the excessive long-lived afterglow (phosphorescence) that is characteristic of CsI[Tl] and other scintillators (\cite{afterglow1}, Fig.\ 3). This is a crucial feature in an application involving the detection of faint scintillation signals, in particular for a detector operated in a location lacking significant overburden. In such a site, the frequent large energy depositions from cosmic-ray interactions ($\sim$1 / cm$^{2}$ min) can lead to an intolerable continuum of delayed afterglow emissions in other scintillators, resulting in an increase of the effective threshold. 

\item Several other practical advantages exist: CsI[Na] exhibits a high light yield ($\sim$45 photons/keVee at 662 keVee \cite{lyield}). With a light emission peaking at 420 nm, it features the best match to the response curve of bialkali photocathodes of any scintillator. This results in almost twice the photoelectron yield of CsI[Tl], for similar energy depositions. CsI[Na] exhibits a short scintillation decay time ($\tau_{f}\sim0.6 
~\mu$s, $\tau_{f}\sim1~\mu$s for CsI[Tl]), suitable for the timing characteristics of the neutrino emission at a spallation source. It is a rugged detector material, and exhibits a negligible light yield variation in the temperature range 15 C-30 C. It is also relatively inexpensive ($\sim$ \$1,000/kg), permitting an eventual detector mass large enough (few hundreds of kg) to attempt the interesting CENNS physics studies mentioned in Sec.\ 1. 

\end{itemize}

In this paper we describe the development of a 14 kg CsI[Na] detector in a dedicated radiation shield, intended for installation at the SNS during 2014 within the framework of the CSI:SNS collaboration\footnote{Recently renamed as COHERENT collaboration.} \cite{white}. This device is expected to be sufficient for a first measurement of CENNS at the SNS, while providing operational experience towards a larger-mass detector. 

\section{Characterization of the response of CsI[Na] to low-energy nuclear recoils}
\label{}

Nuclear recoils are not as efficient in producing scintillation or ionization as electron recoils of the same energy. This is true of  any detector medium. The ratio between the scintillation yields for these two recoiling species is referred to as ``quenching factor", and is typically of order few percent at low energy. Therefore, a crucial first step before a CENNS measurement is a careful characterization of this quenching factor, in the relevant few keVnr to few tens of keVnr range. Its present measurement for CsI[Na]  uses a dedicated setup described in detail elsewhere \cite{myq2}, employed in that publication for the characterization of NaI[Tl]. The reader is referred to \cite{myq2} for a full description of the method,  which features several improvements with respect to previously-employed techniques. \\

Briefly stated, a small 14.5 cm$^{3}$ CsI[Na] crystal was exposed to the monochromatic 2.2 MeV emissions from a DD neutron generator, capturing the scattered neutrons with a second scintillator capable of neutron/gamma discrimination. The scattering angle is varied by changing the position of this second detector on a goniometric table, scanning in the process a range of nuclear recoil energies (8 keVnr-30 keVnr) in the crystal under test. Separately, the response to electron recoils was explored using Compton scattering from a collimated beam of 356 keV $^{133}$Ba gammas, captured by a germanium detector following their interaction with CsI[Na]. Again, the Compton scattering angle was scanned, leading to a measurement of the response to electron recoils in the range 3 keVee-50 keVee (``ee" makes reference to ``electron equivalent", i.e, ionization energy). This small CsI[Na] detector was obtained from the same source \protect\cite{proteus} as the 2 kg and 14 kg crystals discussed below, using the same sodium dopant concentration. Scintillation was measured using a high-quantum efficiency UBA PMT \cite{myq2}. \\

Fig.\ 4 shows example distributions of light yield for several choices of neutron scattering angle (i.e., nuclear recoil energy). Cs and I recoils are indistinguishable due to their very similar nuclear mass. This figure is equivalent to Fig.\ 8 in \cite{myq2}. The light yield obtained for electron recoils via Compton scattering is also displayed in the inset, similar to Fig.\ 4 in \cite{myq2}. Following \cite{myq2}, the ratio between these yields  is employed to extract the quenching factor in the energy region of interest for a CENNS exploration at the SNS (Fig.\ 5). \\

Similar to the situation described in \cite{myq2,myq3} for NaI[Tl], the low-energy quenching factor found here is considerably smaller than in previous attempts by others \cite{qfkims}. A discussion of systematic effects able to artificially increase low-energy quenching factor measurements is provided in \cite{myq2,myq4}. Several measures designed to avoid these systematics were in place during the present measurement \cite{myq2}. While the adoption of our smaller quenching factor is very conservative vis-a-vis a calculation of CENNS rates at the SNS, a separate measurement using a dedicated monochromatic pulsed neutron beam line at TUNL's tandem accelerator is in progress: preliminary results using NaI[Tl] \cite{private} reproduce the trend towards a monotonically decreasing quenching factor with decreasing energy, found here and in \cite{myq2,myq3}. A confirmation of the same behavior for CsI[Na] and CsI[Tl] would have an important impact on the claimed sensitivity of a recent dark matter search using cesium iodide \cite{newkims}. \\

Fig.\ 6 displays the scintillation decay time observed during present calibrations, for both electron recoils (ER) and nuclear recoils (NR). The evolution with energy evident in the figure has been observed before for CsI[Tl] \cite{qfkims} and NaI[Tl] \cite{nai60,myq2,othernai}. Small differences in decay time have been exploited for the statistical discrimination between NR's and ER's in dark matter searches, once a sufficiently large number of events is gathered \cite{nai60}. In view of Fig.\ 6, it may be possible to apply a similar statistical discrimination approach to eventually demonstrate that POT-coincident events at the SNS do indeed originate dominantly in nuclear recoils. In order to further explore this possibility, two subsets of the NR and ER available data were selected, each containing 1,000 events, under the criterion of having distributions as close as possible in number of photoelectrons (PE) registered per event (Fig.\ 7, inset). Waveforms for these 1,000 events were then added together, paying attention to assigning a zero-time coordinate to the position of the first photoelectron in the waveform\footnote{Also removing from the analysis the artificial initial spike in PMT current generated when performing this action.} (Fig.\ 7). The average low-energy NR and ER waveforms thus obtained are then fitted allowing for a fast and slow scintillation decay component \cite{decaytime} plus a (negligible) constant term. Keeping in mind that 10 PE corresponds here to $\sim$0.55 keVee and to $\sim$12 keVnr, these fits confirm the evolution with energy observed in Fig.\ 6. This indicates that it may indeed be possible to distinguish a group of $\gtrsim$1,000 NR events from its ER equivalent, through a measurement of the fast decay time constant of their ensemble. The planned approach to data-acquisition at the SNS (Sec.\ 5) would allow to test this interesting possibility, by comparing the cumulative of NR-dominated events in the few $\mu$s following POT, to those in the few $\mu$s prior to POT, the latter being representative of ER-dominated steady-state backgrounds not associated to pulsed SNS emissions. However, this test would require a substantial further improvement to the signal-to-background ratio presently achieved (Sec.\ 5). 

\section{Detector and shielding design}
\label{}

In order to assess the steady-state environmental and internal backgrounds affecting a deployment at the SNS, we developed a 2 kg prototype of the 14 kg CsI[Na] CENNS detector. The complete shielding intended for SNS installation was populated with this prototype, and operated in a 6 m.w.e. laboratory at the University of Chicago, an overburden similar to that available at the SNS basement location discussed below. The two detectors are identical in all respects, except for their length. The expectation is that the larger device will exhibit a slightly reduced low-energy background per unit detector mass, due to its enhanced peak-to-Compton ratio.\\

A $\sim$9.9 PE/keVee yield is measured for few hundred keVee gamma interactions in the 2 kg prototype\footnote{This light yield is expected to increase by 30\% at 10 keVee, due to the intrinsic non-linearity characteristic of CsI[Na] \cite{nonlinearity}. We observe an increase of precisely this magnitude in our low-energy Compton scattering calibrations (Fig.\ 4), after accounting for the difference in quantum efficiency between UBA (33\%) and standard bialkali PMTs (22\%) for CsI[Na] scintillation wavelengths.}. Adopting the 22\% quantum efficiency expected from the standard bialkali photocathode employed and the CsI[Na] spectral emission, this corresponds to the nominal $\sim$45 photon/keVee light yield from CsI[Na] \protect\cite{lyield}, indicating an optimal light collection efficiency. Based on existing information \cite{uniformity} and manufacturer data, only a small decrease in overall light collection efficiency is expected from the larger crystal, by limiting its length to 33 cm and using a large active area (11.5 cm diameter) PMT for the readout. This efficiency will be characterized along the long axis of the final crystal using a collimated gamma-ray scan. \\

Figs.\ 8 and 9 show different aspects of detector and shielding. Samples from the crystal boule were screened against internal radioactive contaminants using low-background germanium gamma spectroscopy at SNOLAB, and ICP-MS. The concentration of $^{238}$U, $^{235}$U and $^{232}$Th was below the $\sim$1 ppb sensitivity of the measurements, however showing a presence\footnote{It is worth mentioning that the presence of a small $^{40}$K contamination provides a convenient built-in low-energy calibration reference at 3.2 keVee. This is the binding energy of K-shell electrons in the argon daughter following $^{40}$K electron capture, detectable whenever a 1,461 keV gamma de-excitation escapes the crystal.} of 17$\pm$16 mBq/kg of $^{40}$K. The material contained detectable levels of radioactive Cs and Rb isotopes (26$\pm$2 mBq/kg of $^{134}$Cs, 28$\pm$3 mBq/kg of $^{137}$Cs, and 72 ppb of natural Rb -i.e., 20 ppb $^{87}$Rb-). These are known to be responsible for most of the internal low-energy backgrounds in this scintillator \cite{radiopurekims}. These concentrations were determined to be sufficiently low for the present application (Sec.\ 5). Crystals are wrapped in PFTE expanded-membrane reflector. Low-background Suprasil synthetic silica windows are employed. Epoxies and optical couplants were screened against excessive internal radioactivity. The crystals are encapsulated in electroformed OFHC copper cans custom-grown at the Pacific Northwest National Laboratory (PNNL). A low-background Electron Tubes bialkali 9390UFL PMT, responsible for just 0.5$\pm$0.3 Bq of gamma emissions above 0.1 MeV, is used. A custom-designed OFHC copper structure holds crystal and PMT in place, also providing support for two inches of ancient lead, placed between detector and high-activity electronics in its voltage divider (Fig.\ 8). This ultra-low background (ULB) lead, also present in a 1" innermost layer surrounding the detector, was measured to contain $<$0.02 Bq/kg of $^{210}$Pb using radiochemical extraction and high-sensitivity alpha spectroscopy at PNNL. This inner lead layer results in a negligible low-energy background from  $^{210}$Pb bremsstrahlung \cite{vojtila}. The gamma shielding is completed with contemporary  lead ($\sim$100 Bq $^{210}$Pb/kg), resulting in a minimum of 18 cm of lead surrounding the CsI[Na] crystals in any direction.\\

Shielding against thermal neutrons is present in the form of borated silicone (sides, top) or cadmium sheet (bottom) layers immediately outside of the lead castle. Surrounding these is a set of seven plastic scintillator muon veto panels (sides and top), two inches thick. Each panel is read out by four summed-output 3/4" PMTs, inserted into the panel so as to obtain a compact assembly devoid of protruding light guides. Each set of four PMTs is gain-matched, and their common high-voltage bias is adjusted to obtain an optimal separation between muon interactions and lower-energy signals from environmental gammas, by avoiding veto discriminator triggers on the second. The efficiency of this active shield against events generated by cosmic rays interactions in detector or shielding is measured at $>$99.6\% by monitoring the spectral reduction above 10 MeV, an energy regime dominated by such events.  The triggering rate from a Lecroy 620BL 8-channel discriminator used to provide a common logic output from the ensemble of veto panels is just $\sim$300 Hz for this veto efficiency. This rate is in good agreement with expectations based on the muon flux at sea level, and the surface area of this veto. This modest triggering rate is expected to generate a negligible incidence of false veto coincidences during operation at the SNS. \\

Fig.\ 9 shows this assembly inserted into a steel pipe, which supports additional external polyethylene neutron moderator. This pipe is intended for an eventual installation in a shallow (few meter deep) underground well outside of the SNS building. Prompt fast neutrons with energies of up to several hundred MeV are able to escape the massive shielding monolith surrounding the SNS spallation target, with a time-profile of arrival to the detector that can approximate that for the neutrino emissions. This time profile for different beam-associated backgrounds is illustrated in Fig.\ 10.  Therefore, highly-penetrating hard neutrons able to induce nuclear recoils constitute the main background of concern in a search for CENNS at a spallation source\footnote{Nevertheless, their recoil energy spectrum and time profile should allow for their discrimination at the expense of some neutrino signal loss: the fastest neutrons induce large recoil energies at early times (Fig.\ 10), whereas the most energetic recoils from CENNS are due to $\nu_{e}$ and $\bar{\nu}_{\mu}$ from $\mu$ decay, which can arrive at later times. Preliminary data obtained at the SNS \cite{private2} indicate only a marginal neutron flux $\sim 2~\mu$s following a POT spill, leading to a strong reduction in the backgrounds affecting CENNS from delayed $\nu_{e}$ and $\bar{\nu}_{\mu}$.} \cite{steve,white}. A shallow well is expected to provide sufficient shielding in the direct line-of-sight to the source to remove this concern. An effort to characterize neutron and gamma backgrounds and their time profile at different locations within the SNS facility is underway within the COHERENT collaboration \cite{white}.  Preliminary results using arrays of large liquid scintillator detectors with neutron/gamma discrimination capability indicate that an existing basement corridor, 20-25 m away from the SNS target, may be an ideal emplacement for CENNS detectors. This location provides $\sim$17 meters of concrete and gravel shielding in addition to the target monolith, resulting in a further extinction of the fast neutron flux by some five orders of magnitude \cite{private2}. This may be sufficient to entirely remove their associated backgrounds. Background measurements and simulation efforts continue at the time of this writing.

\section{Background measurements, data acquisition strategy, and prospects for CENNS detection.}
\label{}

Fig.\ 11 displays the steady-state backgrounds measured with the shielded 2 kg crystal under a 6 m.w.e. overburden. No neutron moderator was present during these measurements. This spectrum of energy was obtained using a preamplifier and shaping amplifier, with signals registered by a multi-channel analyzer. This approach is limited: while it is useful to help identify remaining contaminations and to assess the veto efficiency, it can underestimate very small energy depositions due to their limited triggering efficiency, or overestimate them by a lack of discrimination against PMT afterpulses, phosphorescence, etc. \\

In order to properly assess the steady-state background rate down to single photoelectron (SPE) emissions, we mimicked the 60 Hz POT trigger that will be available at the SNS through the use of an electronic pulser. This logic signal provides an external trigger to a National Instruments PCI 5153 digitizer, set to sample the PMT output at 0.5 Gs/s, sufficient to reconstruct SPEs and the integrated current from small few-photoelectron signals. A second available digitizer channel monitors the ``OR" output from the 8-channel discriminator merging the muon veto panel signals, allowing for the rejection of events associated to cosmic ray interactions in detector or shielding. The acquired waveforms are 60 $\mu$s long, of which 50 $\mu$s are pre-trigger information. Steady-state backgrounds not-related to the SNS beam can be measured, and eventually subtracted, by inspecting the PMT current present in the 10 $\mu$s immediately preceding the POT trigger\footnote{A concern must be considered, which is that the pre-POT spectrum might contain energy depositions on the average slightly larger than for the post-POT equivalent, due to the monotonically decreasing phosphorescent yield from earlier energy depositions in the crystal. However, we observe statistically-indistinguishable low-energy ($<$20 PE)  background spectra from the 10 $\mu$s pre- and post-POT trigger intervals during these tests. This absence of a measurable difference over a 10 $\mu$s time shift is to be expected for few-PE signals, in view to the comparatively very long time constants involved in delayed phosphorescent de-excitations able to generate such faint signals (Fig.\ 3).}. The chosen waveform length allows to impose the same data cuts to pre-POT and post-POT signals. A first data cut is based on the presence of an excessive number of SPEs in the 40 $\mu$s preceding the 10 $\mu$s region being inspected. We find that imposing this condition allows for an efficient rejection of low-energy events associated to phosphorescence from a previous large energy deposition, while introducing only a modest  dead time\footnote{The SPE rate observed from this PMT when matched to the 2 kg prototype detector is 6,000 Hz, most of it associated to phosphorescence, i.e., not uniformly distributed in time. The irreducible few-PE background rate in Fig.\ 13 is obtained by imposing the condition of zero PE in the 40 $\mu$s pre-trigger trace, which results in a 75\% live-time. Relaxing this condition to accept one PE in the pre-trigger trace increases this background rate by a factor of two.}. \\

Surviving events involving less than 20 PE in the 10 $\mu$s post-trigger window can be further analyzed by examining the pulse shape of their cumulative (i.e., time-integrated) PMT current distribution. A vast majority of these events display cumulative currents characteristic of a random grouping of their constituent photoelectrons, dissimilar to a radiation-induced scintillation pulse like those shown in Fig.\ 7. This is evidenced by their broad, featureless distribution of cumulative current rise-times, shown in Fig.\ 12. Additional data cuts can be imposed by exploiting this observation. A Monte Carlo simulation of the expected cumulative current from neutrino-induced CENNS events is used to define and optimize these cuts. The simulation employs as input the characteristic fast and slow scintillation components studied in Fig.\ 7. It also includes the measured dispersion in SPE current, as well as the post-trigger time-of-arrival of neutrino signals from the different neutrino flavors, prompt and delayed.  A basic set of pulse-shape cuts can be imposed resulting in 46\% CENNS signal acceptance and 96\% rejection of backgrounds from randomly-grouped PEs. The resulting irreducible low-energy spectrum and accepted CENNS signal rate is shown in Fig.\ 13. We expect a further improved signal-to-background ratio from a more sophisticated treatment of this pulse-shape rejection (e.g., using neural networks \cite{neural} or wavelet analysis \cite{wavelets}).\\

A labVIEW-based DAQ software is optimized to write waveform bundles to file with 100\% throughput at the 60 Hz SNS trigger rate. The software inspects incoming waveforms in real time, not saving to file the $\sim$2/3 of events containing blank (zero PE) traces. The same program periodically compresses raw binary files, discarding uncompressed originals, resulting in an expected manageable 40 Gb/day data stream at the SNS. \\

The steady-state background achieved during these tests is sufficiently small to expect a statistically-significant  evidence for a CENNS-induced excess in the residual between the energy spectra from 10 $\mu$s post-POT and 10 $\mu$s pre-POT time periods, following a reasonable (few year) exposure (Fig.\ 13, inset). A comparison between the presently measured concentration of $^{87}$Rb and $^{134,137}$Cs, and simulations of their contribution \cite{radiopurekims}, indicates that just a $\sim$5\% of the low-energy steady-state background measured in the 2 kg prototype originates from these internal radioisotopes\footnote{Accounting for the duty-cycle imposed by mimicking the SNS trigger and for the measured photoelectron yield per keV, the low-energy background shown in Fig.\ 13 is equivalent to $\sim$600 counts / keV kg day.}. A more exhaustive control of these, as is done in \cite{radiopurekims}, would therefore not lead to a significant enhancement of the CENNS signal-to-background ratio in a future larger CsI[Na] detector array. On the other hand, the fast (0.1-10 MeV) neutron flux in the shallow underground  laboratory where these tests were performed has been measured at $\sim7\cdot10^{-4}$ n/cm$^{2}$s, using a combination of Bonner spectrometers. The exact energy distribution of these neutrons is unknown. Signals from their elastic scattering are expected to accumulate at few PE, due to the limited maximum recoil energy imparted to a heavy nucleus like Cs or I, and the few-percent quenching factor involved. A rough simulation of their contribution   to the spectrum in Fig.\ 13 indicates that a significant further reduction in low-energy background may be achieved through the addition of neutron moderator against these environmental neutrons.\\

Fig.\ 14 shows the CENNS signal together with two beam-coincident neutron-induced signals able to produce competing nuclear recoils in the CsI[Na] crystal. The first is the high-energy neutron component escaping the SNS monolith already mentioned in the previous section, calculated here using a detailed GEANT simulation of neutron production at the SNS target, and its transport to the basement location described above. This simulation requires parallelization, event biasing, and up to $\sim$25,000 CPU-hours in a computer cluster, due to the long distance ($\sim 24$ m) of moderating materials traversed\footnote{Along the direct line of sight from the SNS target to the basement location considered, these are 5.2 m of monolith steel, 3.5 m of high-density concrete, 2 m of regular concrete, 13.2 m of compacted road base stone, and 70 cm of regular concrete \cite{private2}.}. This background calculation will be better informed by the ongoing measurements of neutron flux at the SNS, but it can already be observed to have negligible impact on a CENNS measurement in this basement location, or its equivalent. \\

The second competing signal in Fig.\ 14 arises from the charged-current reaction $^{208}$Pb($\nu_{e},e^{-}$)$^{208}$Bi in the lead shielding around the detector. This reaction is of great interest for supernova neutrino detection in this same energy range \cite{nspectra,halo}, and specifically for the ongoing HALO experiment \cite{halo2}. Its value has not been experimentally assessed yet. The excited $^{208}$Bi daughter is expected to rapidly decay generating up to three neutrons, detectable by the CsI[Na] crystal. The planned CENNS measurement provides an excellent opportunity to parasitically measure this separate neutrino cross-section. Neutrino-induced neutron spallation is expected to play a dominant role in heavy-element nucleosynthesis in supernovae: this additional measurement will contribute experimental information to an area that relies heavily on theoretical models \cite{snspa}.\\

The signal from $^{208}$Pb($\nu_{e},e^{-}$)$^{208}$Bi shown in Fig.\ 14 is calculated as follows\footnote{The cross-section for inelastic neutral-current scattering of $\nu_{x}$ off $^{208}$Pb is one order of magnitude lower than for this charged-current, and is neglected here. Accounting for contributions from all neutrino flavors, the neutral current reaction rate in Pb is expected to add to  the charged current background studied here  by $\sim$15\% \cite{katepr}.}. Neutron production rate in the 2,300 kg of lead shielding surrounding the detector is derived from Table II in \cite{nmultipl}, adjusting for the $\sim1.1\cdot10^{7}$ cm$^{-2}$ s$^{-1}$ $\nu_{e}$ flux expected 25 m away from the SNS target, at the farthest presently considered basement location. This input includes the expected multiplicity in neutron generation\footnote{Table II in \cite{nmultipl} contains two typos: it should read 16350 instead of 1630 (``Pb+1n" line, column ``10 m"), and 2400 instead of 1140 (``Pb+2n" line, column ``20 m") \cite{lazauskas}. These corrections have been taken into account.}. Following \cite{nspectra}, a simple evaporation energy spectrum terminated at 5 MeV is adopted for the initial neutron energies\footnote{An improved assessment of this emission spectrum would be desirable, in view of the factor 2-3 difference between neutrino energies at the SNS and those considered in \cite{nspectra}. However, based on the modest dependence on neutrino energy indicated in \cite{nspectra}, the estimates of Fig.\ 14 are not expected to change by much.}. These neutrons are then transported through a detailed MCNP-Polimi \cite{polimi} geometry of the detector assembly, resulting in 3.6\% of them producing nuclear recoils in the CsI[Na] crystal. For instances of multiple scattering in the crystal, recoil energies are added\footnote{This slightly overestimates the contribution from this process above detector threshold, since the quenching factor (Fig.\ 5) is expected to be monotonically decreasing with decreasing recoil energy.}. It should be kept in mind that an uncertainty by a factor $\sim$3 exists in the calculated cross-section for this charged-current process, specifically for the decay-at-rest $\nu_{e}$ energies expected at the SNS, depending on the nuclear model employed. Even if the $^{208}$Pb($\nu_{e},e^{-}$)$^{208}$Bi cross-section adopted here is one of the largest \cite{nmultipl,xsections,indian}, the CENNS signal is observed to dominate  at energies near the detector threshold (Fig.\ 14). \\

While the spectral shape of the beam-coincident excess should be sufficient to disentangle $^{208}$Pb($\nu_{e},e^{-}$)$^{208}$Bi and CENNS contributions, there are additional strategies available to show the presence of two separate components. For instance, the innermost 1" of ULB lead shielding can be trivially replaced by low-background polyethylene neutron moderator during a second running period\footnote{This action is expected to bring up the observed steady-state low-energy backgrounds by just a few percent, due to the predictable increase in $^{210}$Pb bremsstrahlung \cite{vojtila} and reduction in external gamma shielding \cite{leadthickness}.}. The resulting considerable softening in energy of neutrons from the charged-current reaction in lead, leading to lower energy recoils, can be noticed in Fig.\ 14. The observation of the dissimilar signal rate changes predicted for the $\sim$7-25 keVnr and for the $\gtrsim$25 keVnr energy regions following this inner shield substitution (Fig.\ 14) can be used to confirm the presence of two separate neutrino-induced signals, each dominating in one of these energy ranges. In addition to this, a convenient fast neutron monitor internal to a CsI[Na] detector is provided by the inelastic scattering signal\footnote{This characteristic signal is observed to arise not only from muon-induced neutrons (Fig.\ 11), but also with a very high rate of incidence during DD neutron generator irradiations of CsI[Na] and NaI[Tl] \cite{myq2} (elastic and inelastic scattering cross-sections for 1-10 MeV neutrons in $^{127}$I differ only by a factor of $\sim$2).} at 57.6 keV (Fig.\ 11). Its rate in coincidence with the SNS beam can be used to provide an independent measure of the neutron flux from $^{208}$Pb($\nu_{e},e^{-}$)$^{208}$Bi reaching the CsI[Na] crystal. Another possibility is to measure this $^{208}$Pb($\nu_{e},e^{-}$)$^{208}$Bi channel independently and upfront, by placing a one-liter Bicron-501A liquid scintillator cell within the lead shield in lieu of the CsI(Na) crystal. Exploiting the neutron/gamma discrimination characteristic of this scintillator, it should be possible to identify approximately 40 events per month at the SNS basement from this neutrino-induced reaction, removing the uncertainties presently associated to this background.

\section{Conclusions}
\label{}      

The COHERENT collaboration \cite{white} uses a multi-target approach to CENNS detection at the SNS, in this way maximizing the sensitivity of new investigations in neutrino physics, reachable through this unexploited channel. The collaboration presently contemplates the use of p-type point contact (PPC) germanium detector arrays \cite{ppc}, a two-phase liquid xenon detector \cite{red}, and CsI[Na] detectors like those described here. In addition to these, a recently decommissioned multi-ton array comprising 1,000 NaI[Tl] detectors previously used for national security applications, has recently become available at PNNL.  An interesting use for these crystals, beyond the mentioned CENNS-related studies, would be an attempt to observe the pattern of $\nu_{\mu}$ oscillations into sterile neutrinos within the detector array itself, a possibility reachable for $\Delta m^{2}\gtrsim 7$ eV$^{2}$ \cite{osc}. The characteristics of these NaI[Tl] detectors (background, light yield, quenching factor) are being presently assessed by the collaboration.\\

To summarize, we have demonstrated the potential of CsI[Na] as a suitable target for CENNS studies at a spallation source. Necessary conditions of detector threshold, target mass, and background level, have been shown to be in place for a first measurement of this intriguing mode of neutrino interaction, following a short (few years) exposure with a 14 kg CsI[Na] crystal. It should be possible to extract a first measurement of the $^{208}$Pb($\nu_{e},e^{-}$)$^{208}$Bi cross-section, of interest for future supernova detection, from the same effort. The present feasibility study relies strongly on the availability of a detector location at the SNS with sufficient neutron moderating material in the direct line-of-sight to the target. The COHERENT collaboration is presently performing background studies in a number of detector emplacements potentially meeting this requirement.

\section{Acknowledgements}
\label{}      
                                              
We are indebted to our colleagues from the COHERENT collaboration for many useful exchanges, in particular to P.S. Barbeau, Yu.\ Efremenko and K. Scholberg, and to I. Lawson for facilitating many sample measurements at SNOLAB. N.E.F. was supported by the NNSA Stewardship Science Graduate Fellowship program under grant number DE-FC52-08NA28752. T.W.H. was supported by the Intelligence Community Postdoctoral Research Fellowship Program. Additional support was received from the Kavli Institute for Cosmological Physics at the University of Chicago through grant NSF PHY-1125897, and an endowment from the Kavli Foundation and its founder Fred Kavli. J.I.C. acknowledges the hospitality of the International Institute of Physics at the Universidade Federal do Rio Grande do Norte during the completion of this manuscript. This work was completed in part with resources provided by the University of Chicago Research Computing Center.

\bibliographystyle{elsarticle-num}
\bibliography{<your-bib-database>}

\newpage

\begin{figure}
\includegraphics[width=10cm]{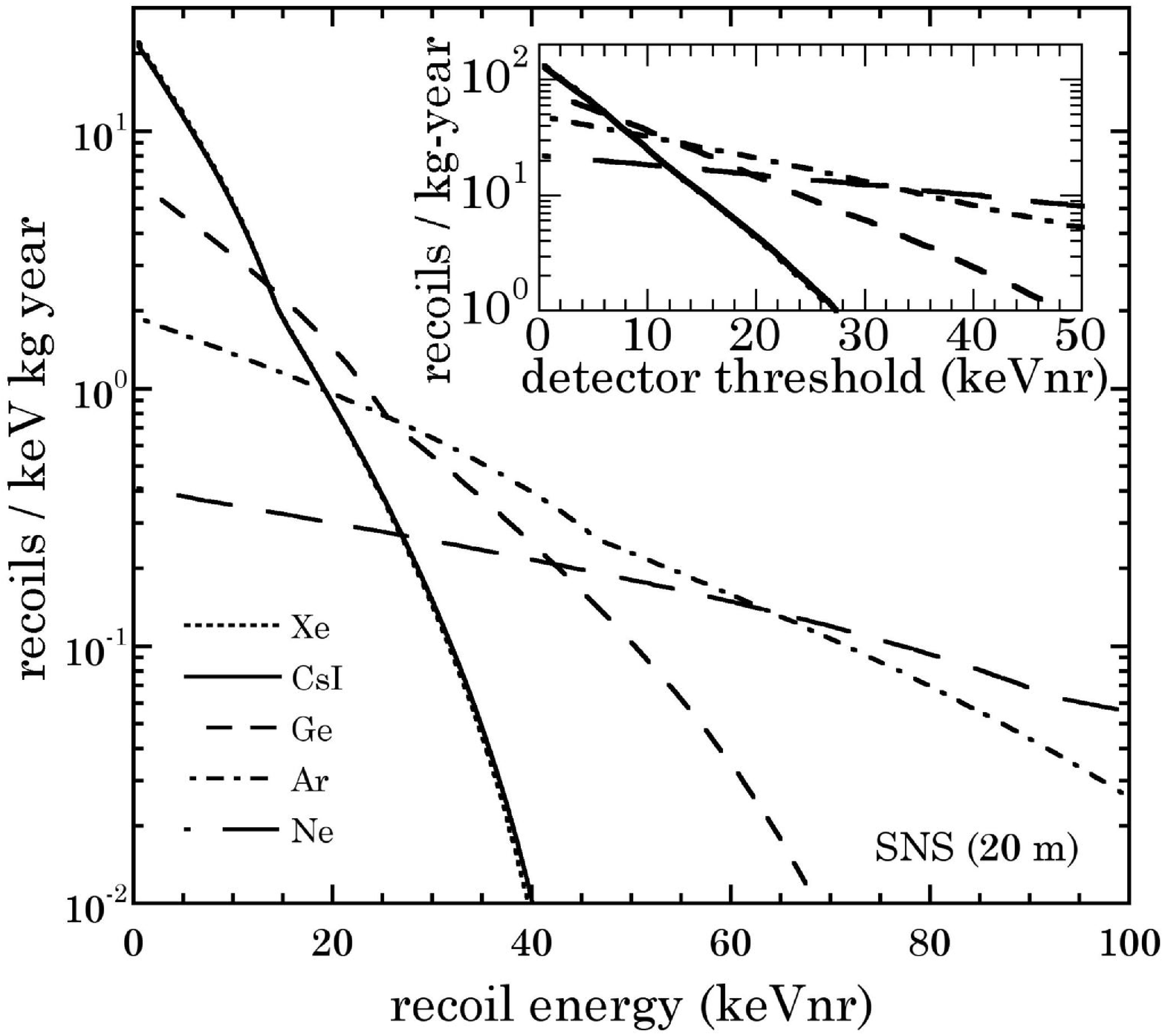}
\caption{Comparison between nuclear recoil energy spectra from different CENNS detector materials at the SNS. The calculation follows the formalism of \protect\cite{drukier} and includes a nuclear form factor as in \protect\cite{pfsmith}. Approximately 0.13 $\nu$ are produced per proton on target, per flavor. For present SNS running conditions, the neutrino flux at 20 m from the SNS interaction point is $\sim1.7\cdot10^{7}$ cm$^{-2}$ s$^{-1}$, per flavor. {\it Inset:} CENNS interaction rate integrated above detector threshold. The tradeoff  between coherent enhancement to the cross-section and recoil energy is evident in this figure: heavy targets are favored, but only as long as a sufficiently-low threshold can be achieved. 
}
\end{figure}

\begin{figure}
\includegraphics[width=10cm]{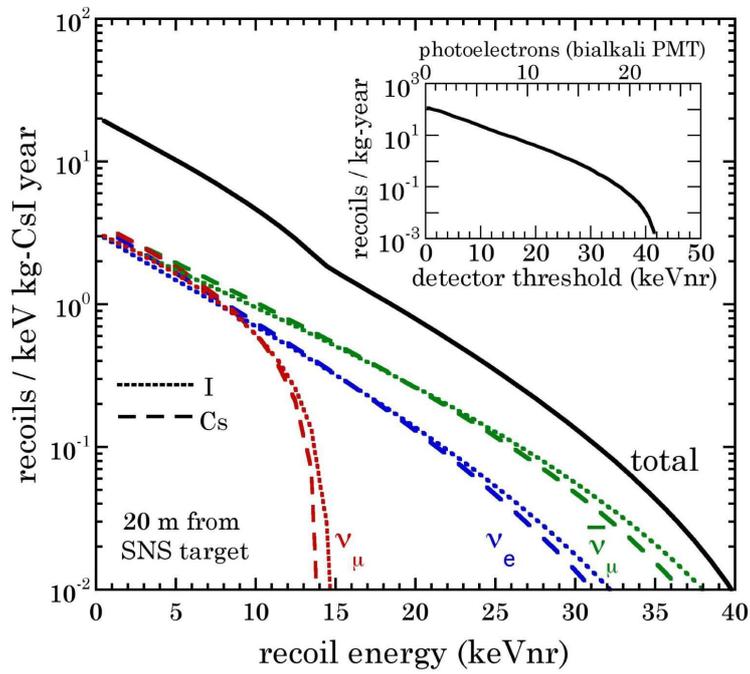}
\caption{Response of CsI[Na] to CENNS at the SNS, segregated by recoiling species and neutrino flavor. The sodium dopant is present with a fractional mass of just $10^{-4}-10^{-5}$, playing no significant role as a target. Cs and I respond almost identically to a given neutrino
flavor, simplifying the task of understanding the detector response. 
{\it Inset:} Integrated rate above detector threshold in nuclear recoil energy (keVnr). This energy scale is translated into number of detected photoelectrons (PEs) in a conventional bialkali PMT through the quenching factor measured in Sec.\ 3 and the  light yield in the 2 kg prototype at few keVee (Sec.\ 4). Test runs performed in similar conditions to those expected at the SNS show that a threshold of 4 PE ($\sim$7 keVnr) is reachable with sufficient signal-to-background ratio (Sec.\ 5).
}
\end{figure}

\begin{figure}
\includegraphics[width=10cm]{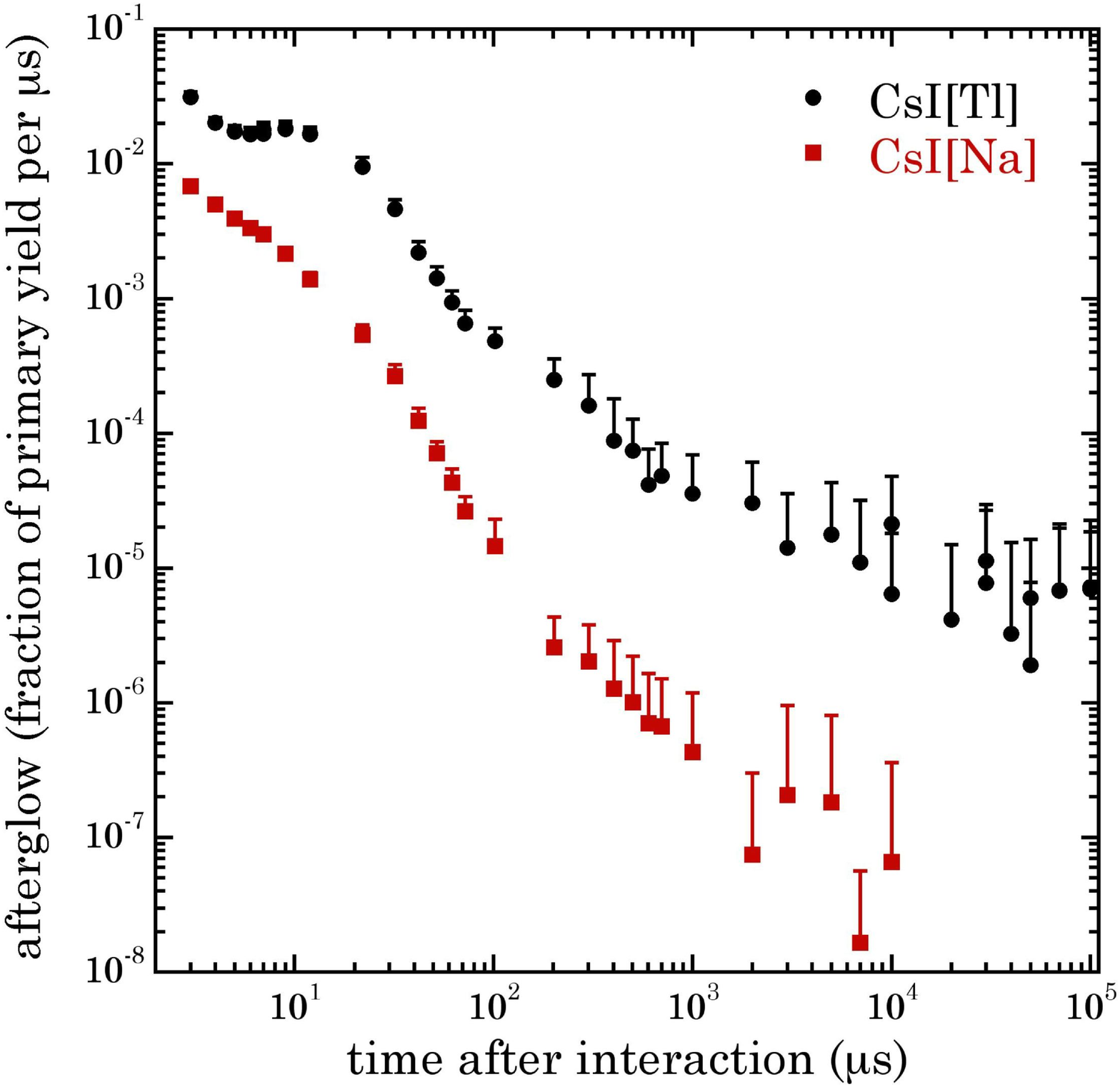}
\caption{Afterglow (phosphorescence) in CsI[Na] and CsI[Tl] scintillators. This figure shows the integrated PMT current detected over one $\mu$s, at times following gamma energy depositions of 1.5 ($\pm 0.2$) MeV, as a fraction of the same for the initial primary scintillation. A combination of single-channel analyzer and a gate generator was used to trigger a waveform digitizer in this measurement. Only positive statistical error bars are shown for clarity. Each data point is the average of 100 measurements. Identical small crystals of each material were used, obtained from the same source \protect\cite{proteus} as the 2 kg and 14 kg CsI[Na] detectors discussed in the text. CsI[Tl] is notorious for its excessive afterglow. This limits its use in several applications \protect\cite{afterglow2}, including a CENNS measurement in a site lacking significant overburden (see text). As a reference, a cosmic muon loses 6-10 MeV per cm of cesium iodide traversed \protect\cite{cmuon}, generating a primary scintillation in the range of 40,000-60,000 photons/MeV \protect\cite{lyield}.
}
\end{figure}

\begin{figure}
\includegraphics[width=10cm]{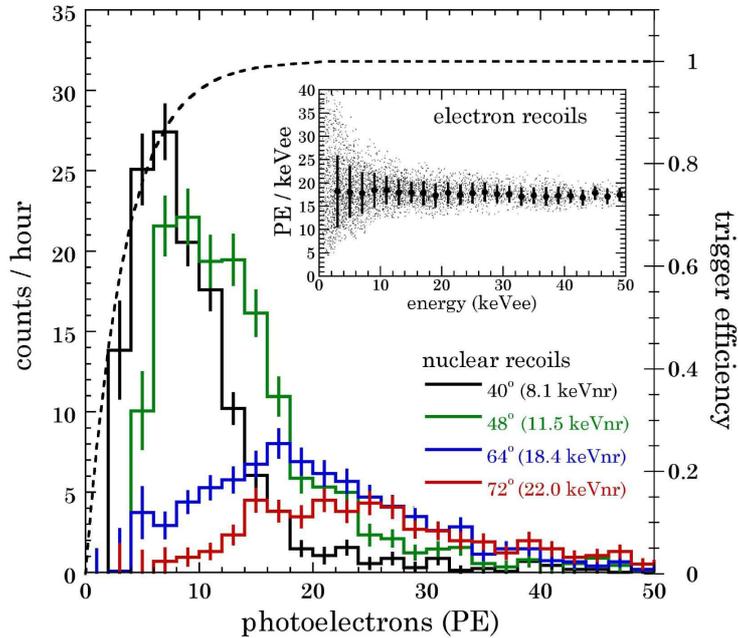}
\caption{Example distributions of photoelectron yield from Cs and I nuclear recoils in CsI[Na], labeled by neutron scattering angle and mean recoil energy. A dotted line represents the triggering efficiency, used to correct these distributions prior to quenching factor determination \protect\cite{myq2}. {\it Inset:} photoelectron yield from low-energy electron recoils measured using Compton scattering  \protect\cite{myq2}.  Taking a calculated 33\% quantum efficiency for the CsI[Na] spectral emission and the UBA photocathode response curve, this is in excellent agreement with the $\sim$45 photon/keVee light yield found at 662 keVee in \protect\cite{lyield} (the 30\% non-linearity in the gamma response of CsI[Na] \protect\cite{nonlinearity} is included in this comparison). Dots correspond to the mean value computed for 2 keVee bins. The small (few percent) quenching factor for low energy nuclear recoils in CsI[Na] is immediately evident when comparing the photoelectron yield for nuclear and electron recoils of the same energy (see text).}
\end{figure}

\begin{figure}
\includegraphics[width=10cm]{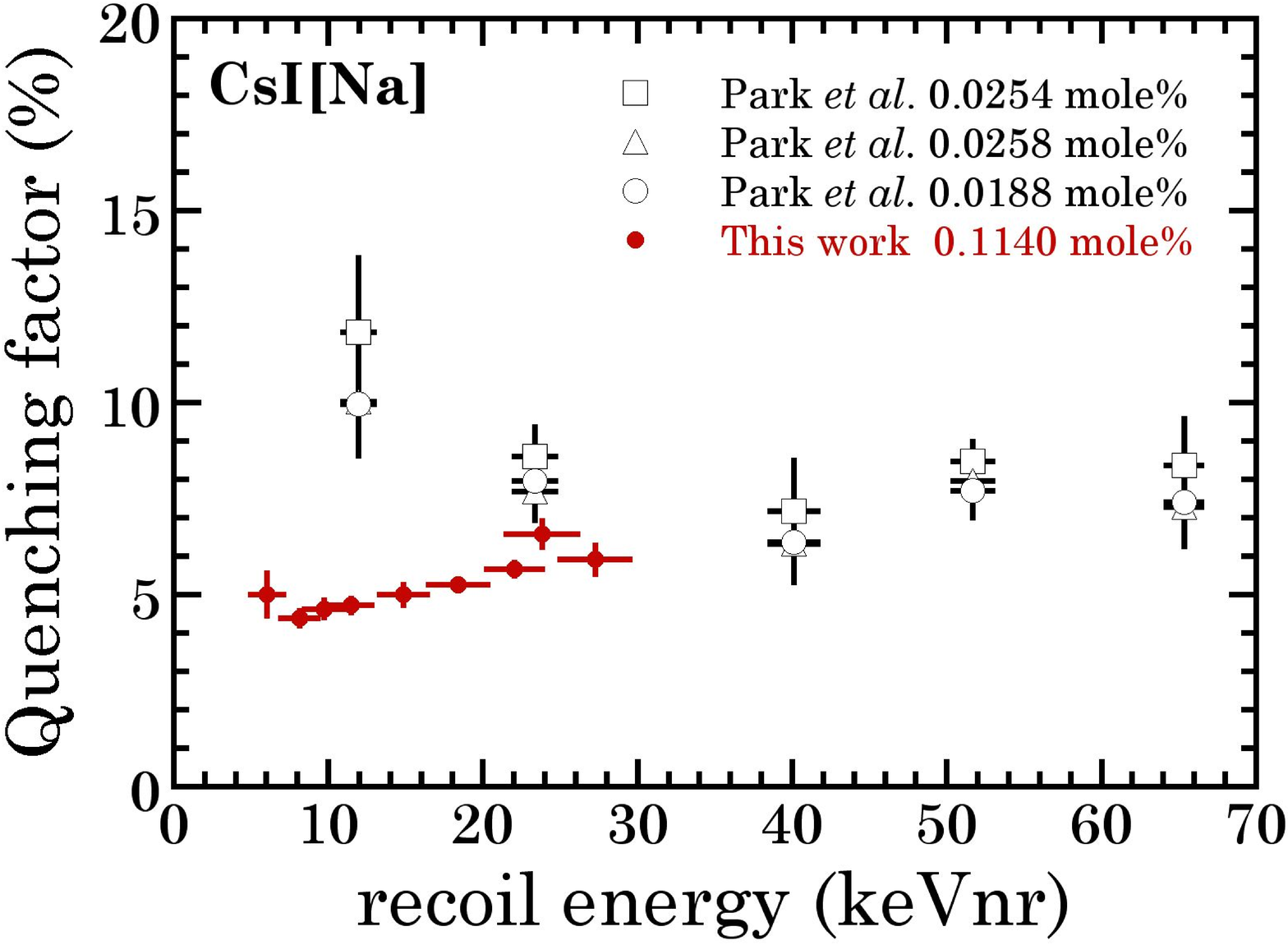}
\caption{Quenching factor for low-energy nuclear recoils in CsI[Na], measured using the setup and method described in \protect\cite{myq2}. The recoil energies probed in this study span the range of interest for a CENNS measurement at the SNS. Horizontal errors bars correspond to the dispersion in the recoil energies probed for each angular orientation of the scattered neutron detector, obtained via MCNP-Polimi simulation \protect\cite{myq2}. Vertical error bars correspond to the uncertainty of the mean in log-normal best fits to distributions like those shown in Fig.\ 4.  A conservative quenching factor of 4.5\% is adopted here for all calculations of CENNS signal rates. See text for a discussion on the origin of the divergence with respect to previous low-energy measurements by Park {\it et al.} \protect\cite{qfkims}. The concentration of Na dopant is indicated in the figure, for each measurement.}
\end{figure}

\begin{figure}
\includegraphics[width=10cm]{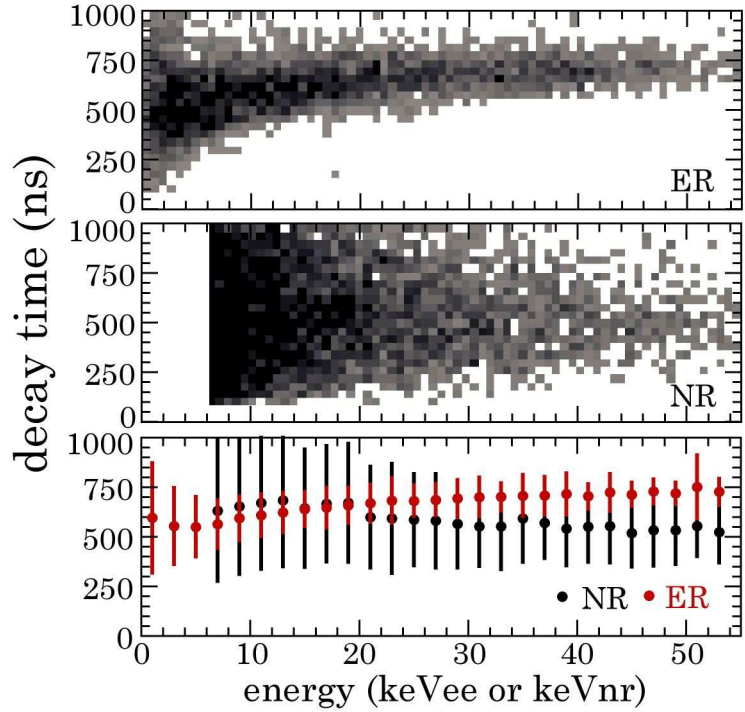}
\caption{Scintillation decay time for nuclear and electron recoils in CsI[Na], approximated using a single exponential fit. The bottom panel shows the mean and dispersion in the grayscale intensity plots, as in Fig.\ 10 of \protect\cite{myq2}. This dispersion is similar for electron and nuclear recoils producing comparable light yields, i.e., once the quenching factor is accounted for.}
\end{figure}

\begin{figure}
\includegraphics[width=10cm]{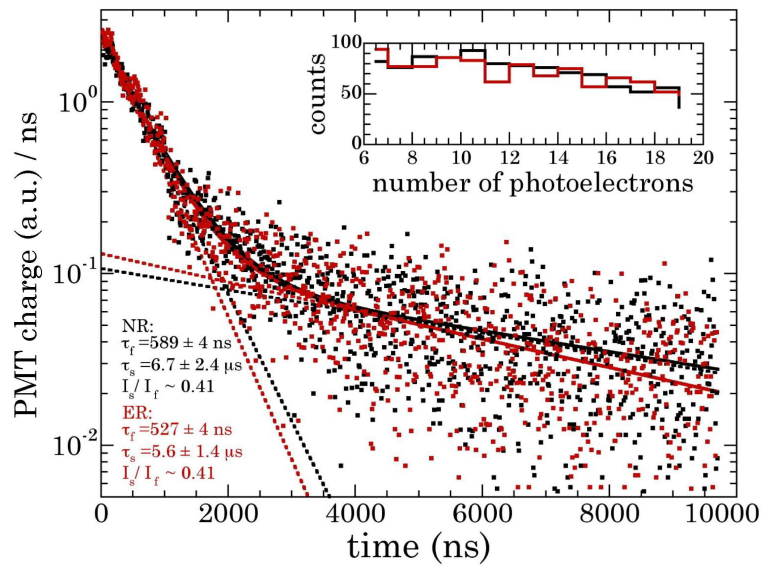}
\caption{Decomposition of the scintillation decay time of CsI[Na] into fast and slow components \protect\cite{decaytime}, for an ensemble of co-added 1,000 nuclear recoils and 1,000 electron recoils (see text). Constant terms allowed by the fit are 1.4$\cdot10^{-5}$ and 1.4$\cdot10^{-3}$ for ER and NR, respectively.  Only one in ten sampled points in the waveforms is displayed, for clarity. NR data are from the two angular runs that provide the closest PE distribution per event to ER data (inset), so as to allow for an unbiased comparison. A statistically-significant $\sim$60 ns difference in the fast decay component ($\tau_{f}$) is noticeable, similar to one previously found for NaI[Tl] recoils at slightly higher energies \protect\cite{nai60}. The ratio of the PMT current induced by the slow and fast components is indicated in the figure. No evidence of a very fast ($\sim$15 ns) decay component for nuclear recoils is found. This is observed under alpha irradiation (and claimed to extend to low-energy NR's) in \protect\cite{fastchinese}.}
\end{figure}

\begin{figure}
\includegraphics[width=14cm]{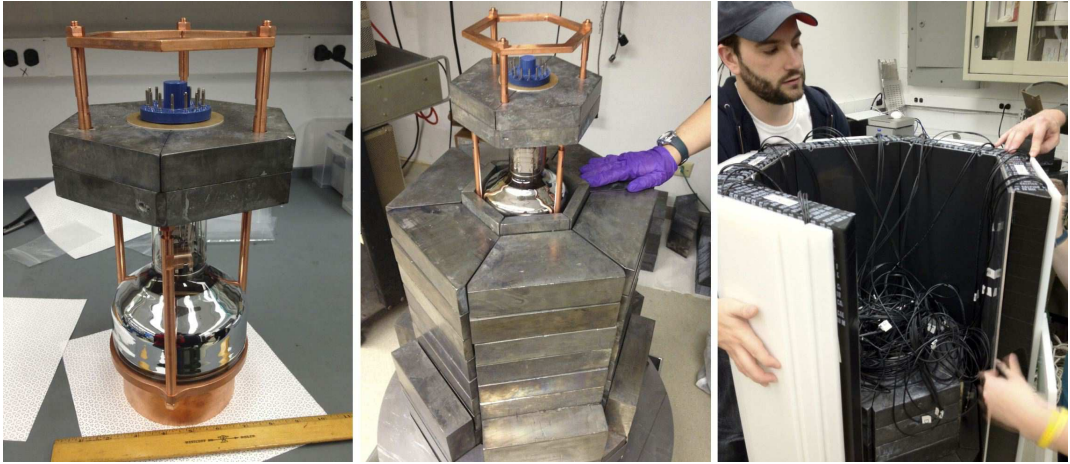}
\caption{Assembly of a 2 kg prototype CsI[Na] crystal within the  shield intended for SNS installation. The detector features selected salts screened against radioactive contaminants, OFHC and electroformed copper parts, a low-background Electron Tubes 9390UFL PMT, archeological-quality lead shielding ($<$0.02 Bq/kg $^{210}$Pb), a 2"-thick muon veto, fast neutron moderator, and thermal neutron absorber. {\it Left:} crystal and PMT. Insulated flying leads allow for the shielding of emissions from high-activity electronic components in the PMT voltage divider behind 2" of ULB Pb. {Center:} crystal inserted in the unfinished lead castle. The innermost 1" layer of ULB lead is visible. A minimum of 18 cm of Pb surrounds the detector in any direction in the completed shield.  {Right:} test of side muon veto panels. Their edges are beveled to obtain a seamless coverage. A round disc completes the top of this active veto.}
\end{figure}

\begin{figure}
\includegraphics[width=10cm]{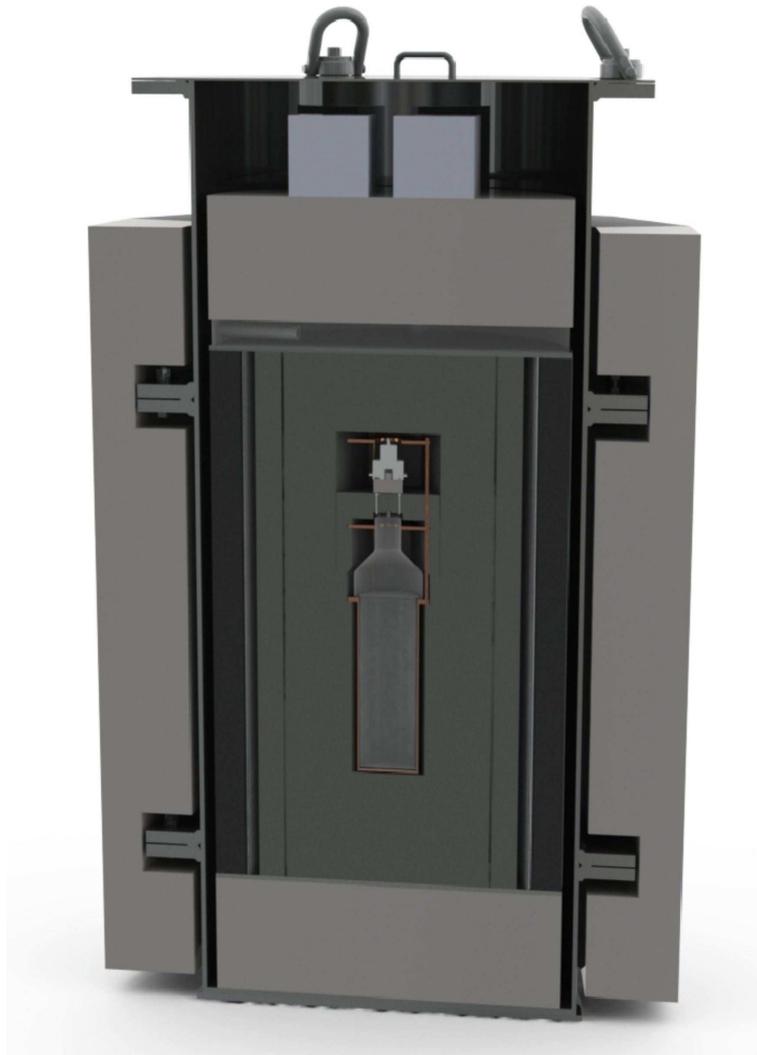}
\caption{Rendering of the 14 kg detector in a full shield. An outermost 4$\pi$ layer of 8"-10" polyethylene neutron moderator is visible. The design includes a steel pipe for eventual lowering of the assembly into a shallow well outside of the SNS building, a measure against backgrounds induced by high-energy neutron leakage from the SNS target monolith. An existing basement location at the SNS may however provide sufficient shielding in the direct line-of-sight to the target (see text).}
\end{figure}

\begin{figure}
\includegraphics[width=14cm]{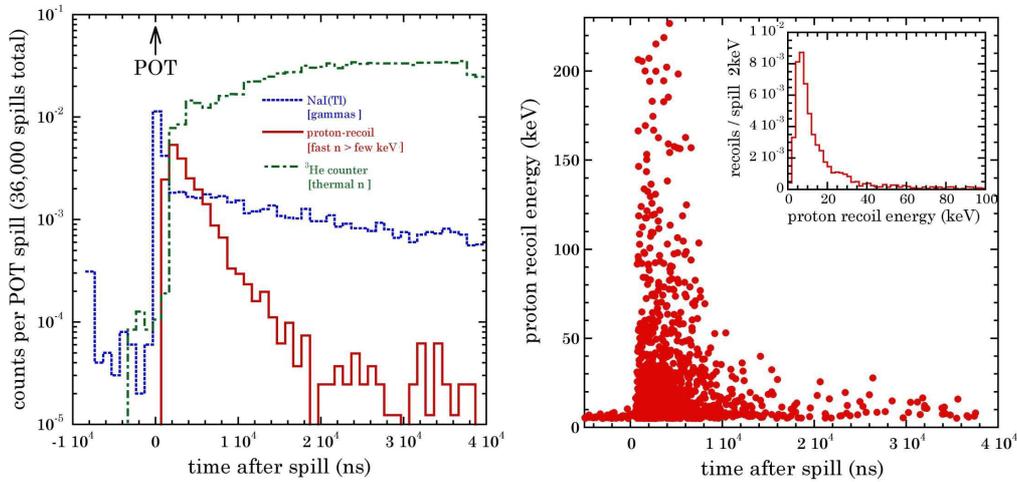}
\caption{{\it Left:} Time-profile for beam-associated backgrounds at a spallation source. Data were collected using an array of three detectors at the IPNS, a lower-intensity precursor of the SNS, now decommissioned at Argonne National Laboratory. A proton-recoil (``Benjamin" \protect\cite{benjamin}) neutron spectrometer was used to isolate the contribution from fast neutrons. Thermal neutrons exhibit a longer build-up and latency due to the finite time necessary for fast neutron moderation and thermalization. The gamma component was measured using a NaI[Tl] scintillator. Measurements were performed with unshielded detectors at the IPNS bay, $\sim$20 m away from target.  {\it Right:} Inverse correlation between measured proton recoil energy and time after POT, expected from the dependence of neutron time-of-flight on neutron energy. The differences in time profile and energy distribution of nuclear recoils from beam-related neutrons and those from neutrino emissions can be exploited  to identify a CENNS signal (see text).}
\end{figure}

\begin{figure}
\includegraphics[width=10cm]{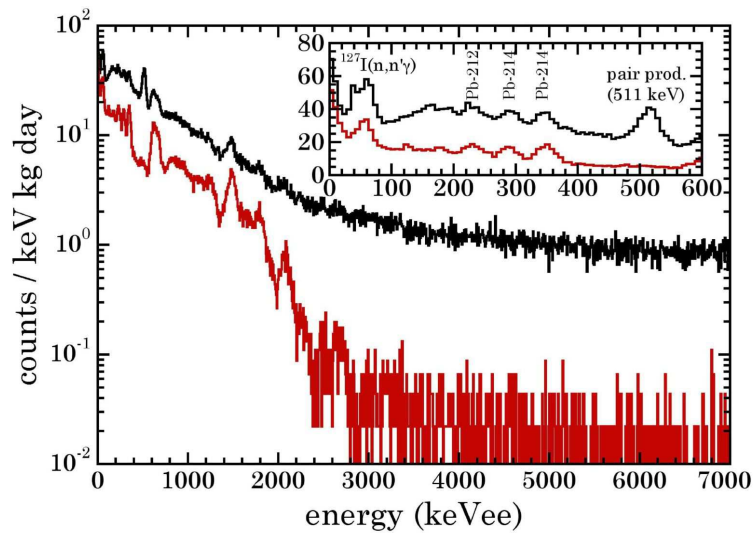}
\caption{High-energy spectra from the 2 kg prototype in a full shield (with the exception of neutron moderator), under 6 m.w.e., an overburden similar to that expected at the SNS. The result of rejecting veto-coincident events is shown by bottom spectra. The inset displays the reduction in annihilation radiation (511 keV) and $^{127}$I(n,n'$\gamma$) signals associated to pair and neutron production, respectively, from cosmic muons interacting preferentially in the lead shield. Environmental neutrons independently contribute to $^{127}$I(n,n'$\gamma$). This peak from inelastic scattering at 57.6 keV \protect\cite{myq2} can be used as a convenient internal monitor of the fast neutron flux reaching the detector (see text).}
\end{figure}

\begin{figure}
\includegraphics[width=10cm]{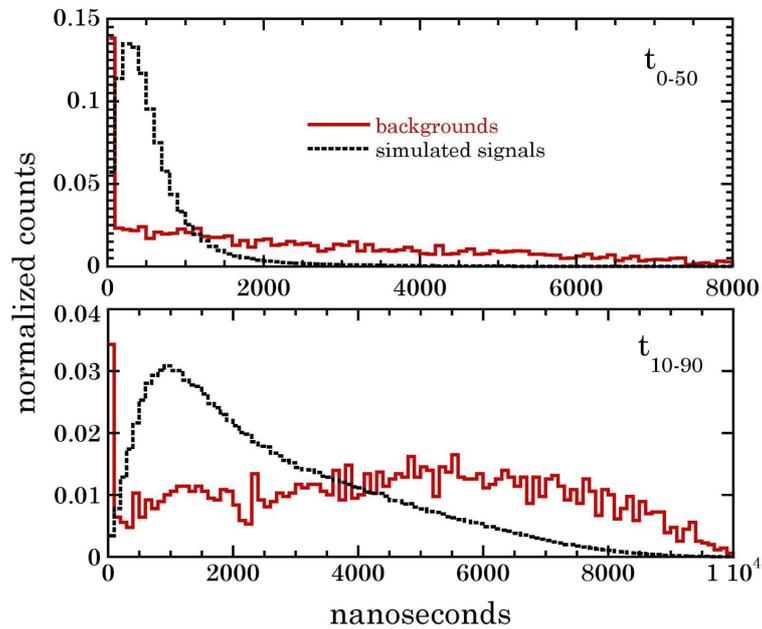}
\caption{Rise time distributions for the cumulative (i.e., time-integrated) PMT current observed for low-energy backgrounds in the 2 kg CsI[Na] prototype, and simulated CENNS pulses based on the scintillation profiles shown in Fig.\ 7 and CENNS nuclear recoil spectrum in Fig.\ 2. Data and simulations are restricted to events in the interval 4-20 PE, and a 10 $\mu$s window following POT (see text). The sub-index (0,10,50,90) indicates the fraction (percent) of the maximum integrated current used for the determination of the rise-time (e.g., t$_{10-90}$ makes reference to the rise-time associated to a current increase from 10\% to 90\% of this maximum). Low-energy backgrounds are dominated by random groupings of photoelectrons originating in PMT dark-current and CsI[Na] phosphorescence, not associated to an actual instantaneous low-energy deposition in the crystal. Event-by-event cuts based on these dissimilar distributions can be used to generate a considerable improvement to the low-energy signal to background ratio (see text).}
\end{figure}

\begin{figure}
\includegraphics[width=10cm]{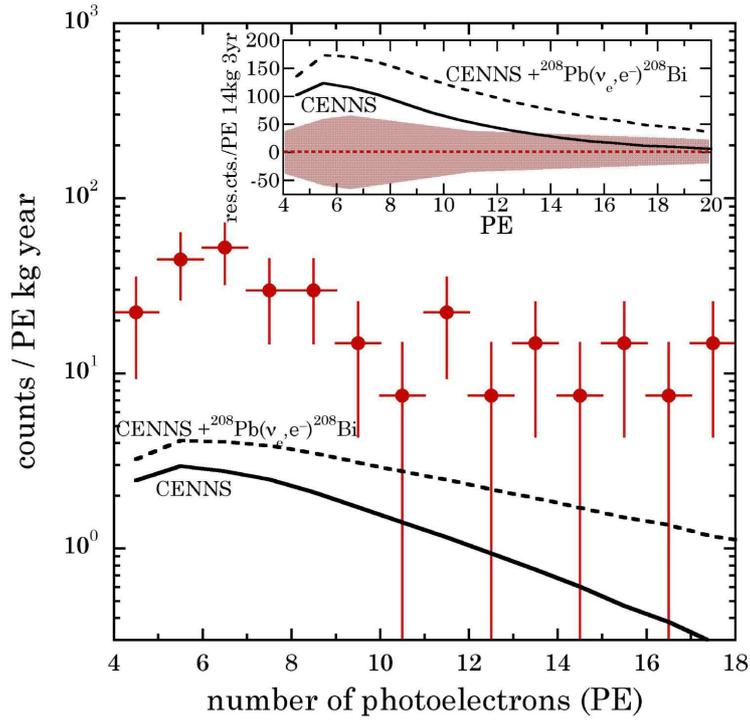}
\caption{Low-energy steady-state background (data points) obtained by mimicking the SNS trigger with the 2 kg CsI[Na] prototype during a 23.3 day run. An additional reduction in this background may be possible through the use of moderator against environmental neutrons (see text). Overlapped is the expected CENNS signal 20 m from the SNS target, with and without the contribution from $^{208}$Pb($\nu_{e},e^{-}$)$^{208}$Bi neutrons, following data cuts described in the main text. The measured 4.5\% quenching factor (Sec.\ 3) and photoelectron yield from a bialkali PMT (Sec.\ 4) are employed to generate this CENNS prediction. {\it Inset:} The subtraction of steady-state background events from the 10 $\mu$s window prior to the POT trigger is expected to generate significant evidence for CENNS in the residual energy spectrum, in this figure shown following 3 years of exposure with a 14 kg CsI[Na] crystal sited 20 m from the SNS. The red band indicates the $\pm 1 \sigma$ statistical fluctuations around a zero residual in the absence of neutrino-induced signals or beam-coincident backgrounds, based on the presently obtained steady-state background level. The expected neutrino-induced excess is also shown. A similar excess from CENNS should appear in an alternative representation showing the time of arrival of signals following POT, due to the characteristic time-profile induced by prompt and delayed neutrino emissions (Sec.\ 1). Steady-state backgrounds are expected to be uniformly distributed in time.}
\end{figure}

\begin{figure}
\includegraphics[width=10cm]{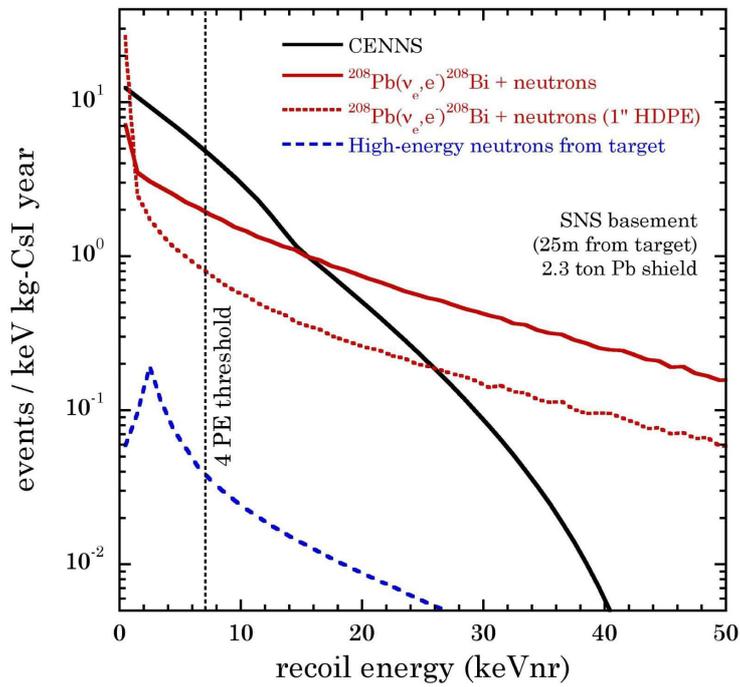}
\caption{Beam-associated neutrino signals and backgrounds expected in the 14 kg CsI[Na] detector at the SNS. Nuclear recoils induced by high-energy neutrons escaping the SNS target monolith can be sufficiently suppressed in a basement location discussed in the main text (the simulation shown here includes an electron recoil component from gamma emission following neutron thermalization).  The ratio of CENNS events above threshold to those induced by the neutron emission following $^{208}$Pb($\nu_{e},e^{-}$)$^{208}$Bi is 1.2:1. This ratio becomes 3.3:1 when the innermost 1" of Pb in the shielding is substituted by high-density polyethylene (HDPE). The possibility exists for a convincing simultaneous measurement of both neutrino cross-sections (see text).}
\end{figure}

\end{document}